\documentclass[manuscript]{aastex}

\usepackage{graphicx}   
\usepackage{amsmath}    
\usepackage{xcolor}     

\shorttitle{BGO}
\shortauthors{Short, Lane \& Fields}

\begin{document}


\title{The Burke-Gaffney Observatory:  A fully roboticized remote-access observatory
with a low resolution spectrograph}


\author{C. Ian Short}
\affil{Department of Astronomy \& Physics and Institute for Computational Astrophysics, Saint Mary's University,
    Halifax, NS, Canada, B3H 3C3}
\email{ian.short@smu.ca}

\author{David J. Lane}
\affil{Department of Astronomy \& Physics and Institute for Computational Astrophysics, Saint Mary's University,
    Halifax, NS, Canada, B3H 3C3}
\email{dave.lane@smu.ca}

\author{Tiffany Fields}
\affil{Department of Astronomy \& Physics and Institute for Computational Astrophysics, Saint Mary's University,
    Halifax, NS, Canada, B3H 3C3}
\email{tiffany.fields@smu.ca}




\begin{abstract}

We describe the current state of the Burke-Gaffney Observatory (BGO) at Saint Mary's University -
a unique fully roboticized remote-access observatory that allows students to carry out imaging, photometry,
and spectroscopy projects remotely from anywhere in the world via a web browser or social media.  Stellar spectroscopy
is available with the ALPY 600 low resolution grism spectrograph equipped with a CCD detector. 
We describe our custom CCD spectroscopy reduction procedure written in the Python programming
language and demonstrate the quality of fits of synthetic spectra computed with the ChromaStarServer 
(CSS) code to BGO spectra.  The facility along with the accompanying Python BGO 
spectroscopy reduction package and the CSS spectrum synthesis code provide an accessible
means for students anywhere to carry our projects at the undergraduate honours
level.  BGO web pages for potential observers are at the site: observatory.smu.ca/bgo-useme.
All codes are available from the OpenStars www site:
openstars.smu.ca/
 
\end{abstract}



\section{Introduction}

The Burke-Gaffney Observatory (BGO, lat. $+44\, 37\, 50$, long. $-63\, 34\, 52$) at Saint Mary's 
University (SMU) underwent a significant refurbishment in 2013 and now consists of a 0.6 m (24'')
$f/6.5$ Planewave telescope (model CDK24) with a corrected Dall-Kirkham optical configuration.  It is  
equipped with an $f/4$ model PF0035 ALPY 600 grism spectrograph from Shelyak Instruments.  
This optical combination involves an inexpensive lightweight spectrograph of low $f$ value paired with a 
telescope of larger $f$ value and is a compromise between optical matching and affordability.  
We operate the spectrograph
with a slit width of 23 $\mu$m corresponding to a spectral resolving power, $R$, of $\sim 600$, equivalent to
a spectral resolution element of $\Delta v \sim 500$ km s$^{-1}$ at $\lambda\sim 6000$ \AA.  
The camera is a model Atik 314L+ equipped with a Sony ICX 205AL sensor CCD with $1391\times 1039$ imaging pixels of size $6.45\times 6.45$ $\mu$m. 
The setup provides a reciprocal linear dispersion, ${\Delta\lambda /\Delta x}$, of $\sim 420$ \AA\, mm$^{-1}$ and a
spectral range, $\Delta\lambda$, of $\sim 3750$ \AA, effectively covering the entire visible band from $\sim 4000$ to $\sim 7000$ \AA~ after accounting for edge effects.

\paragraph{}
The BGO is a fully automatic, roboticized facility, and imaging and photometric observing requests are handled in
queued observing mode.  
The acquisition of spectra is a much more complex process than direct imaging, and approved observers 
submit a request for a synchronous observing session 
via email or by social media and 
carry out a session remotely via a web browser and the BGO observing portal.  
This process requires low-level access to multiple software applications to control the telescope, 
CCD camera, calibration light sources, and to position targets precisely at the spectrograph slit. 
Manual guiding is often required when the target is faint.

\paragraph{}
The BGO observing portal provides the following functions:
1) Telescope pointing via the Earth-Centred-Universe (ECU) telescope
control and planetarium application and accompanying automatic dome rotation via
the custom dome connection software;  2) A large object database via ECU; 
3) A continuously updated view from the spectrograph guide camera for slit monitoring;
4) Acquisition and immediate display of snapshots of the field of view and of the 
spectrum for preliminary inspection;
5) Science exposures of the spectrum.

\paragraph{}
Our experience is that for bright stars ($V < 7$) the diurnal tracking is very stable and that once the 
star is manually centered on the slit the pointing does not require manual correction.  The ''seeing'' at 
the BGO site is significant ($> 1''$) and much of the light falls outside the slit.  Advantages are that
for bright stars
this provides more uniform illumination across the slit and a significant distribution of light in
the cross-dispersion direction.

\begin{figure}
\includegraphics[width=\columnwidth]{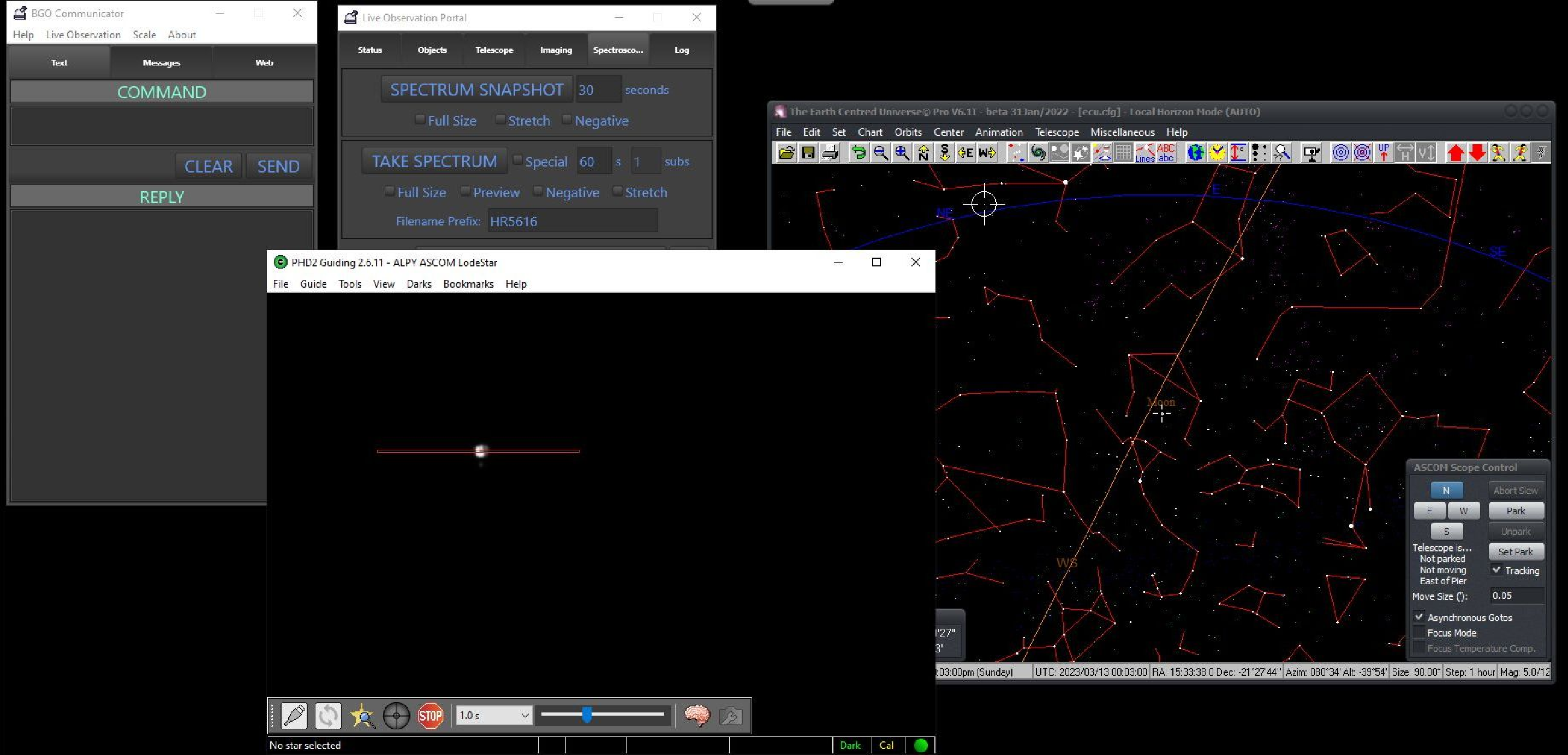}
\caption{HR 5616:  Screenshot of the robotic BGO observing portal during a spectroscopy
session including the ECU planetarium and telescope control interface, and the view from
the spectrograph guide camera.  
  \label{PHD2}
}
\end{figure}

\begin{figure}
\includegraphics[width=\columnwidth]{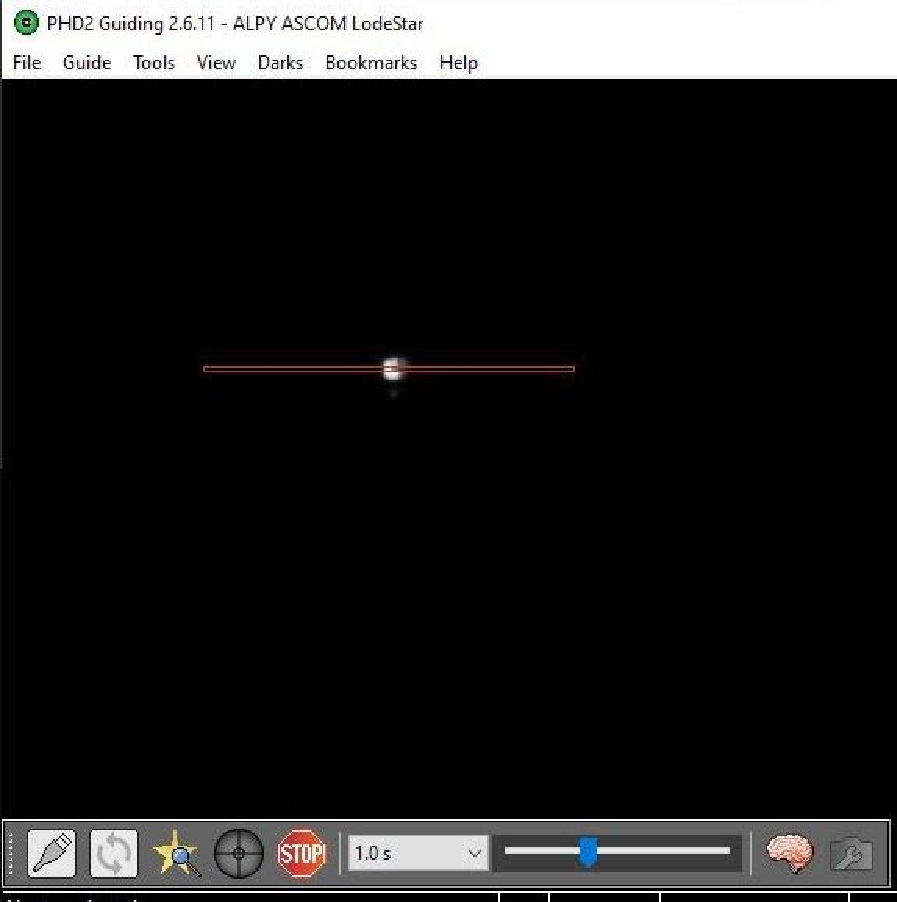}
\caption{The same as Fig. \ref{PHD2}, but showing only the view from
the spectrograph guide camera.
The red rectangle is the projection of the 23 $\mu$m slit onto the sky.
  \label{PHD2Only}
}
\end{figure}

\begin{figure}
\includegraphics[width=\columnwidth]{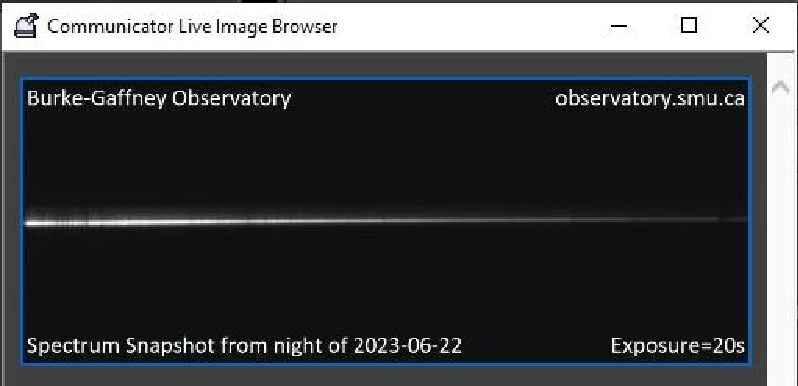}
\caption{Same as Fig. \ref{PHD2}, but showing only a 20 s ''snapshot''
of the spectral image taken with the ALPY 600 spectrograph and the Atik 314L+
imaging camera and Sony ICX 205AL CCD.
  \label{snapOnly}
}
\end{figure}

\paragraph{}
Fig. \ref{PHD2} shows a screen shot of the
BGO observing portal during a spectroscopy session on 22 June 2023 by the author with the 
program star HR5616
($\psi$ Boo, K2 III, $V=4.54$ \citep{hoffleit}), 
including the ECU planetarium and telescope control panel and the live monitoring view from
the spectrograph slit-plane guide camera.  Fig. \ref{PHD2Only} shows an enlarged version of the 
view from the spectrograph guide camera and Fig. \ref{snapOnly} shows a 20 s snapshot of the 
spectral image taken with the Atik 314L+ science camera.  Potential observers, especially students at all 
levels, are encouraged to visit the BGO Robotic Telescope page at observatory.smu.ca/bgo-useme to learn
how to become an approved observer and for a primer on interacting with the BGO remotely.

\section{Observations}

\begin{figure}
\includegraphics[width=\columnwidth]{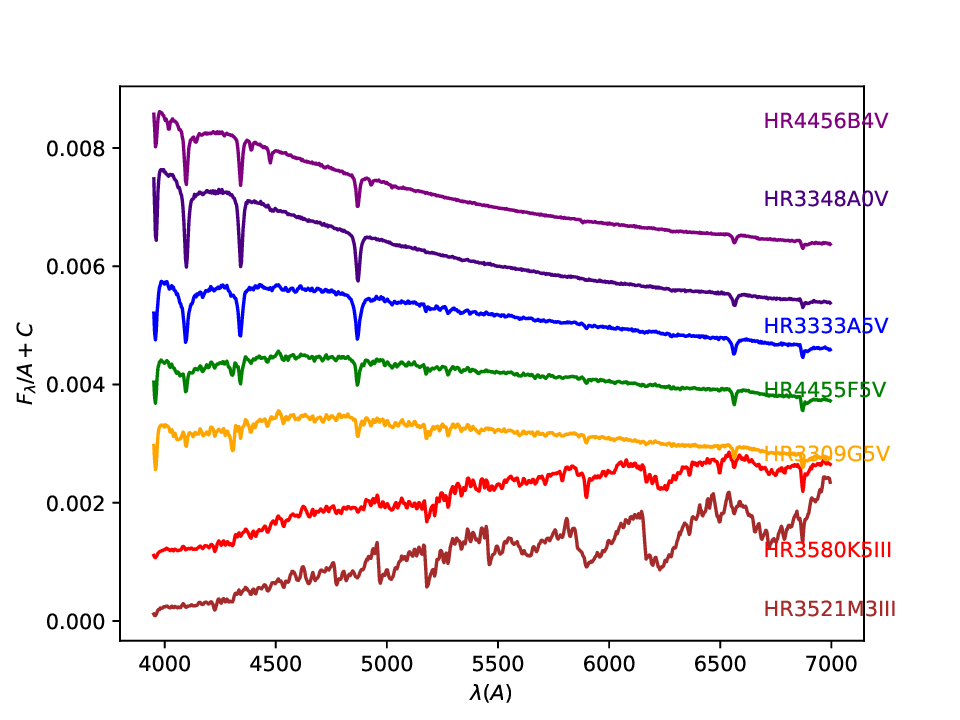}
\caption{Set of seven commissioning spectra for the seven stars listed in Tab. \ref{obslog}
for the ALPY 600 spectrograph at the BGO. 
  \label{DLSequence}
}
\end{figure}

Table \ref{obslog} and Fig. \ref{DLSequence} present a set of commissioning spectra of seven 
bright stars acquired on 
15 April 2021 by one of the co-authors who was the BGO Director
and Astronomy Technician at the time (Lane).  The set includes six luminosity class V
stars spanning the range of spectral class from K5 to B4 and one luminosity class III star of
spectral class M3.  

\begin{table}
\caption{Commissioning stars observed in April 2021 with the ALPY 600 spectrograph.  All
stellar data are those of \citet{hoffleit}}
\label{obslog}
\begin{tabular}{lrrlr}
\tableline
	Designation & $V$ & $B-V$ & Sp. Type & Exp. time (s) \\
\tableline
HR3521  & 6.23 & +1.62  & M3 III & 60 \\
HR3580  & 6.44 & -      & K5 III & 120\\
HR3309  & 6.32 & +0.62  & G5 V   & 120\\
HR4455  & 5.77 & +0.46  & F5 V   & 120\\
HR3333  & 5.95 & +0.19  & A5 V   & 60\\
HR3348  & 6.18 & -0.03  & A0 V   & 60\\
HR4456  & 5.95 & -0.16  & B4 V   & 120\\
\tableline
\end{tabular}
\end{table}

\section{Reduction procedure}

To gain experience with the instrument, and to establish a local reduction procedure that SMU students can modify, we have
chosen to develop an independent procedure in the Python programming language.  In what follows we take CCD pixel columns 
(fixed $x$-coordinate) to run in the cross-dispersion direction, and CCD pixel rows (fixed $y$-coordinate) to run along 
the dispersion axis.  Therefore $x$ ranges from 1 to 1391 and $y$ ranges from 1 to 1039. 

 1) $\lambda$-calibration: 
 We identify nine spectral lines of known wavelength, $\lambda_{\rm n}$ ($n = 0, 1,... 8$), in the emission spectrum of an Ar-Ne lamp, measure their 
 centroid pixel column positions, $x_{\rm n}$, and determine the best fitting coefficients, $a_{\rm m}$ ($m = 0, 1, 2)$, of a $2^{\rm nd}$ 
 order polynomial for the 
 measured $\lambda_{\rm n}(x_{\rm n})$ relation using the numpy \citep{numpy} polyfit() procedure.  We then 
 generate a $1D$ $\lambda$ array of length $1391$ elements, $x$, with the expression  

\begin{equation}
  \lambda(x) = a_2x^2 + a_1x + a_0
\end{equation}

2) Background subtraction: 
 The software package supplied with the ALPY 600 spectrograph includes a post-processing pipeline that automatically applies flat-field division, and bias- and dark-current-subtraction
 corrections.  We extract the value of the 'pedestal'' that is added by the analog-to-digital (ADU) conversion 
 from the FITS header and subtract its value from each pixel.
We have found that this procedure yields images in which there is still an approximately flat background of $\le 10$ counts 
 per pixel across
 the chip in addition to the spectral image, which we infer includes scattered light. 
 We fit a $2D$ function corresponding to a $0^{\rm th}$ order polynomial in both the pixel column, $x$, and pixel row, $y$, dimensions 
 ({\it ie.} a flat background) to two rectangular samples regions on either side of the spectral image and away from the edges of the 
 chip consisting of $1371$ columns ($10 < x < 1381$) and $40$ rows (($10 < y < 50$) and ($989 < y < 1029$)) by finding the average 
 residual count per pixel among the two sample regions, and then subtract the average from every pixel.  This yields a spectral image
 with a residual count of $\sim 0$ counts per pixel.

 3) Spectrum location: 
 The position of the spectral image on the chip, as indicated by the approximate row ($y$ value) where the cross-dispersion
 profile is brightest, depends on the distribution of the stellar point-spread-function (PSF) along the slit, and
 varies significantly from exposure to exposure.
 We automatically locate the row of maximum counts, $y^{\rm max}_{\rm n}(x_{\rm n})$, in three representative columns of $x_{\rm n}$ value 
 of 100, 695 (mid-chip) and 1291 and fit a $1^{\rm st}$ order polynomial to the $y^{\rm max}_{\rm n}(x_{\rm n})$ relation to 
 trace the contour of greatest brightness across the chip.  We find that the dispersion axis is tilted by $\sim 0.5^{\rm o}$ with respect
 to the CCD rows, corresponding to a $\Delta y$ value of $\sim 12$ rows over the $1391$ columns.

\begin{figure}
\includegraphics[width=\columnwidth]{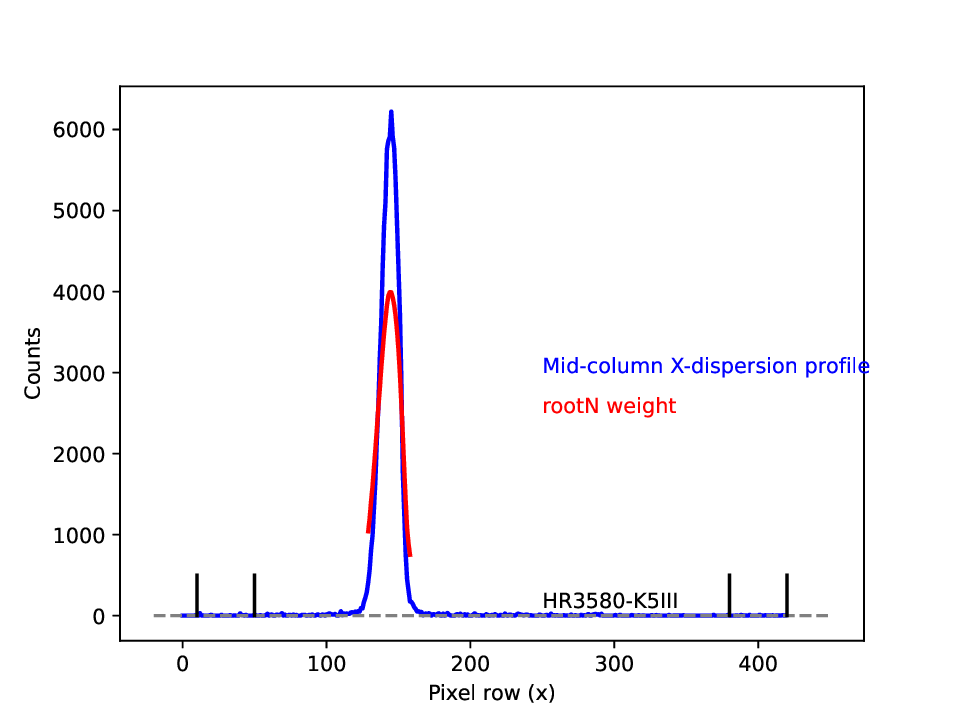}
\caption{Observed cross dispersion profile averaged over 20 columns at chip center
($x = 686$ to $706$, blue line) and the derived root-N weight profile that was used to 
weight the rows ($y$ values) when extracting the 1$D$ spectrum (Step 4, red line).
The weight profile peaks at a value of $\sim 1$ and has been arbitrarily re-scaled for 
visibility.
Also shown are the row ($y$-value) ranges of the regions used to fit
the $0^{\rm th}$ order background signal (Step 2). 
  \label{crossweight}
}
\end{figure}

4) Model cross-dispersion weight profile:  We form a $1D$ normalized smooth cross-dispersion weight profile by 
computing row-wise average counts over the middle 20 columns 
($x = 686$ to $706$) on the chip, taking the square root, and dividing each of the 20 elements in the profile by the 
square root of the maximum number of counts among
the included pixels.  We truncate the wings of the profile at row ($y$) values of the centroid $y$ value $\pm 15$, so 
that the $1D$ profile has 30 elements.  Fig. \ref{crossweight} shows the average observed cross dispersion profile and the
derived weight profile, arbitrarily re-scaled for visibility, for the case of our observation of HR 3580 (K5 III). 

5) Spectrum formation:  We form a $1D$ spectrum by summing the counts in rows ($y$ values) within the range of $\pm 15$ of the row where the
cross-dispersion profile is brightest in each column ($x$ value), weighted by our model cross-dispersion profile 
(see Step 4)).  This root-weighting gives higher weight to pixels near the centre of the cross-dispersion profile
that have higher signal-to-noise ($S/N$) while still taking advantage of the signal in the cross-dispersion wings.

6) Continuum rectification:  Rectification of broadband high $\Delta\lambda /\Delta x$ spectra of late-type
stars is challenging because there are no regions that can be obviously identified as continuum regions.  
Our goal is to develop a procedure that automatically normalizes the spectrum over most of the visible band 
for spectral classes $B$ to $K$, for which the spectrum is not too affected by deep broad TiO bands or by emission lines.
Normalizing the spectrum takes place in three steps:

a) Instrumental response:  An initial very approximate correction is made by dividing the spectrum by a smooth
instrumental response function that was provided by the BGO staff.  
This correction greatly reduces the slope of the spectrum along the dispersion axis. 
Then a $0^{\rm th}$ order normalization is made by dividing every element by the maximum value of this 
quotient spectrum.

b) Fit to truncated {\it mean} binned counts:  We avoid edge effects in our continuum fitting by omitting 
a chosen number of columns at both ends of the spectrum.  We form mean binned counts by choosing a number of bins
and assigning the {\it mean} number of counts in each bin, $f^1_{\rm n}$ to the central pixel ($x_{\rm n}$-value) of that bin.  
We then fit a polynomial to the $f^1_{\rm n}(x_{\rm n})$ relation with the numpy.polyfit() procedure, use the 
best-fit coefficients to generate the corresponding normalizing function $f^1(x)$ and divide the spectrum from Step 6 a) 
element-wise by $f^1(x)$.   

c) Fit to truncated {\it maximum} binned counts:  For late-type stars, Step 6 b) can produce a spectrum with
peak counts significantly greater than unity in local regions throughout the spectrum.  We then follow a similar truncation and
binning procedure as Step 6 b) except we fit a second polynomial, 
$f^2_{\rm n}(x_{\rm n})$ to the {\it maximum} number of counts
among the pixels in each bin, $f^2_{\rm n}$, and divide the spectrum from Step 6 b) by the corresponding
normalizing function $f^2(x)$.

\paragraph{}

\begin{figure}
\includegraphics[width=\columnwidth]{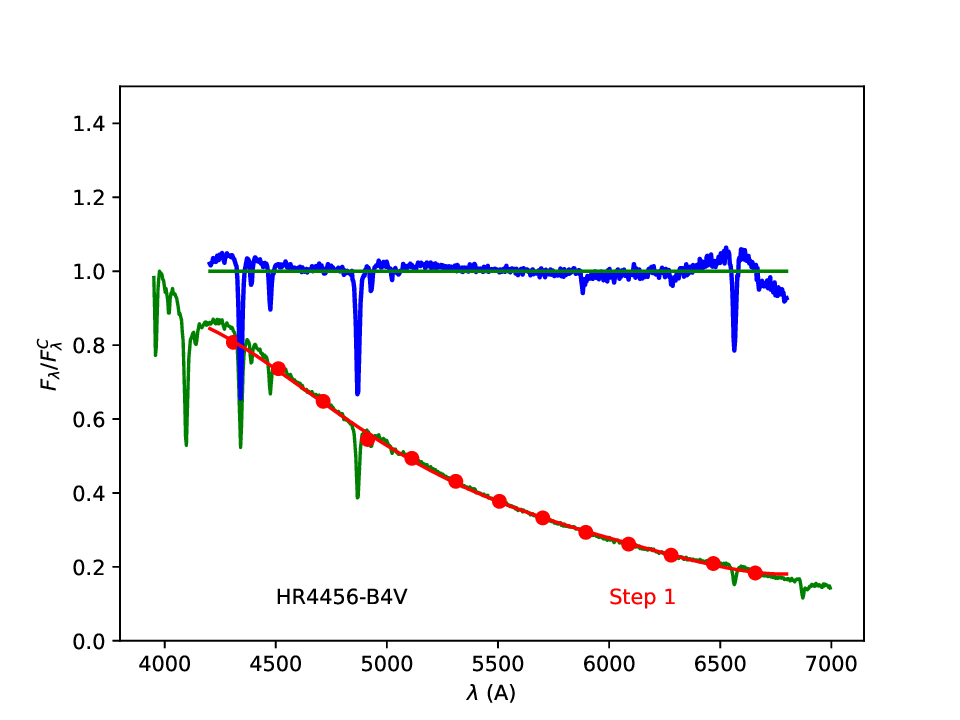}
\caption{Step b) of our general procedure for automatic continuum rectification of broadband spectra
(Step 6) for B4 V star HR 4456.  Green line:
BGO ALPY 600 spectrum after division by an instrumental response function and a $0^{\rm th}$-order
normalization to unity, $F^0(x)$;  Red points:  {\it Mean} binned counts;  Red line:  Polynomial fit
to mean counts points, $f^1(x)$;  Blue line:  Spectrum with refined normalization, $F^1(x) = F^0(x)/f^1(x)$.
Horizontal straight line:  $F(x)$ of unity for reference.
  \label{B4NormB}
}
\end{figure}

\begin{figure}
\includegraphics[width=\columnwidth]{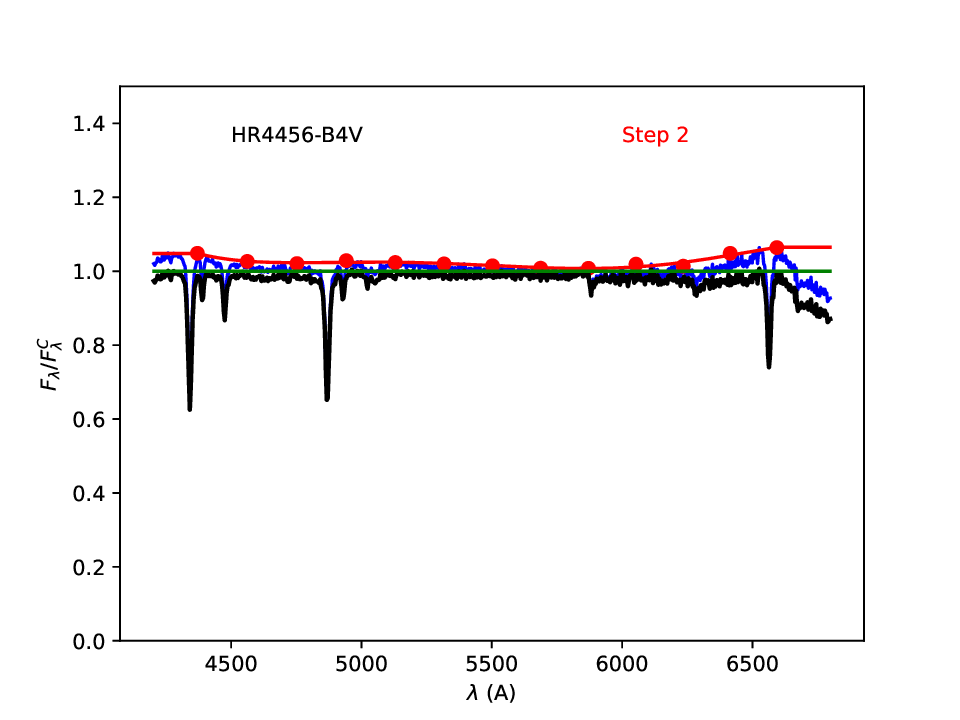}
\caption{Step c) of our general procedure for automatic continuum rectification of broadband spectra for the 
star of Fig. \ref{B4NormB}.  Blue line:
Output from Step b), $F^1(x)$ (see Fig. \ref{B4NormB});
Red points:  {\it Maximum} binned counts;  Red line:  Polynomial fit
to maximum counts points, $f^2(x)$;  Black line:  Spectrum with final normalization, $F^2(x) = F^1(x)/f^2(x)$.
Horizontal straight line:  $F(x)$ of unity for reference.
  \label{B4NormC}
}
\end{figure}

\begin{figure}
\includegraphics[width=\columnwidth]{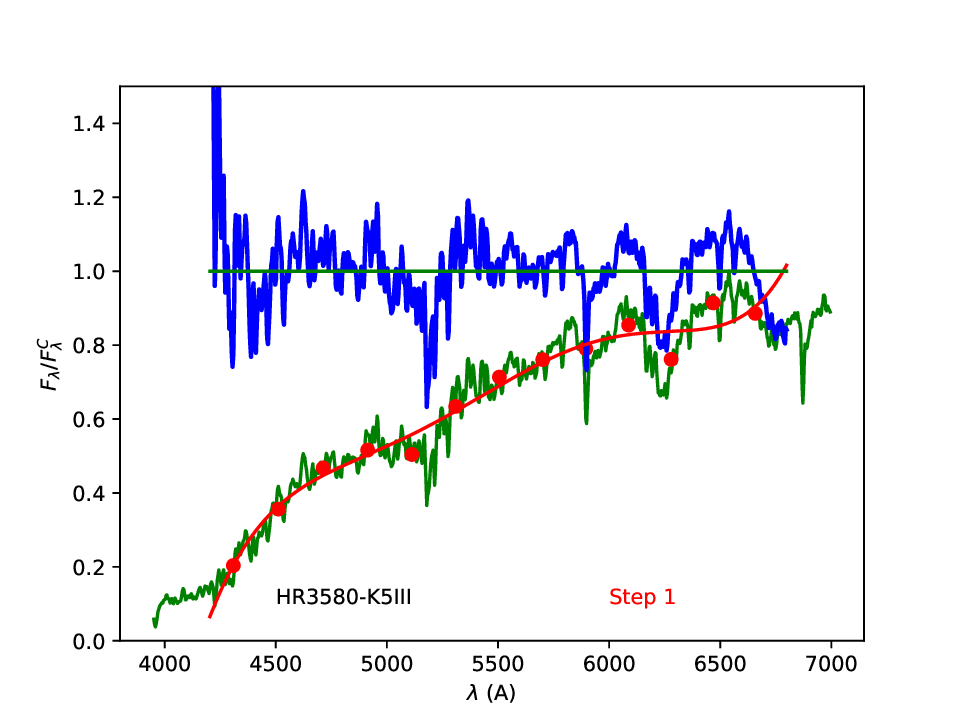}
\caption{Same as Fig. \ref{B4NormB}, except for the K5 III star HR 3580. 
  \label{K5NormB}
}
\end{figure}

\begin{figure}
\includegraphics[width=\columnwidth]{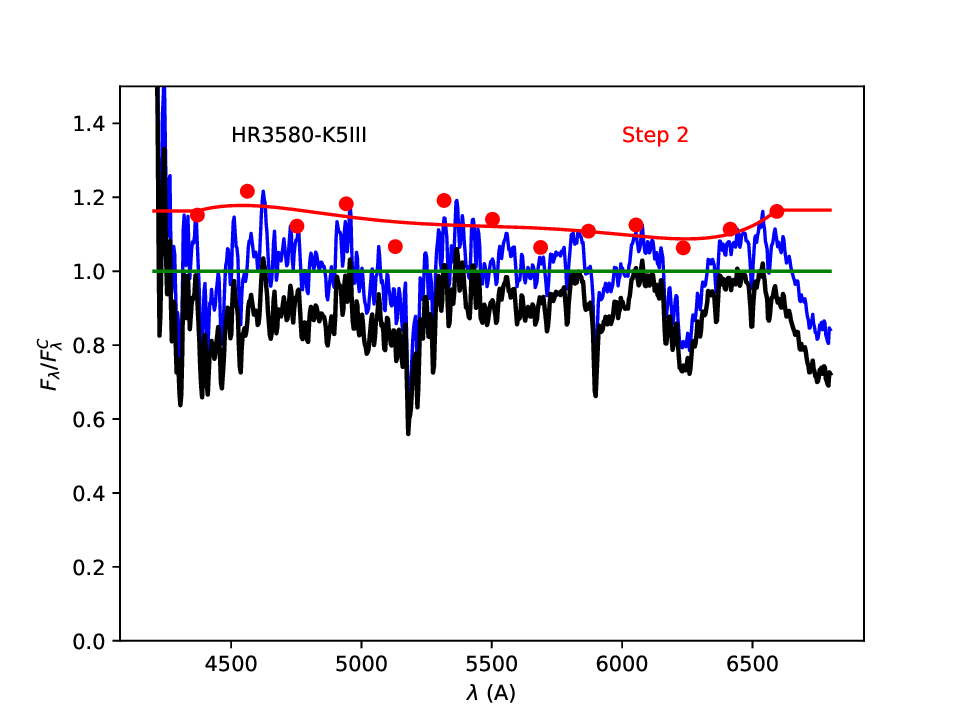}
\caption{Same as Fig. \ref{B4NormC}, except for the K5 III star HR 3580. 
  \label{K5NormC}
}
\end{figure}

Our experience is that we can reliably produce automatically approximately continuum-rectified spectra over much of the central
region of the spectrum for a wide range of spectral classes, including class $K$, by choosing 13 bins for Step 6 b), 
corresponding to $\sim 200$ \AA\, per bin, and fitting with a $5^{\rm th}$ order polynomial,  
and 13 bins for Step 6 c) and a $5^{\rm th}$ order polynomial.  We find we can achieve reasonable normalization
in a spectral region truncated to the range $\lambda$4300 to 6800 \AA.
Figs. \ref{B4NormB} and \ref{B4NormC} show the binned counts, fitting function, and partially re-normalized spectrum
for Steps 6 b) and c) for the case of a B4 V star (HR4456) and Figs. \ref{K5NormB} and \ref{K5NormC} show the 
same for the more challenging case of a K5 III star (HR3580).

\section{Comparison of BGO and model spectra}

\begin{figure}
\includegraphics[width=\columnwidth]{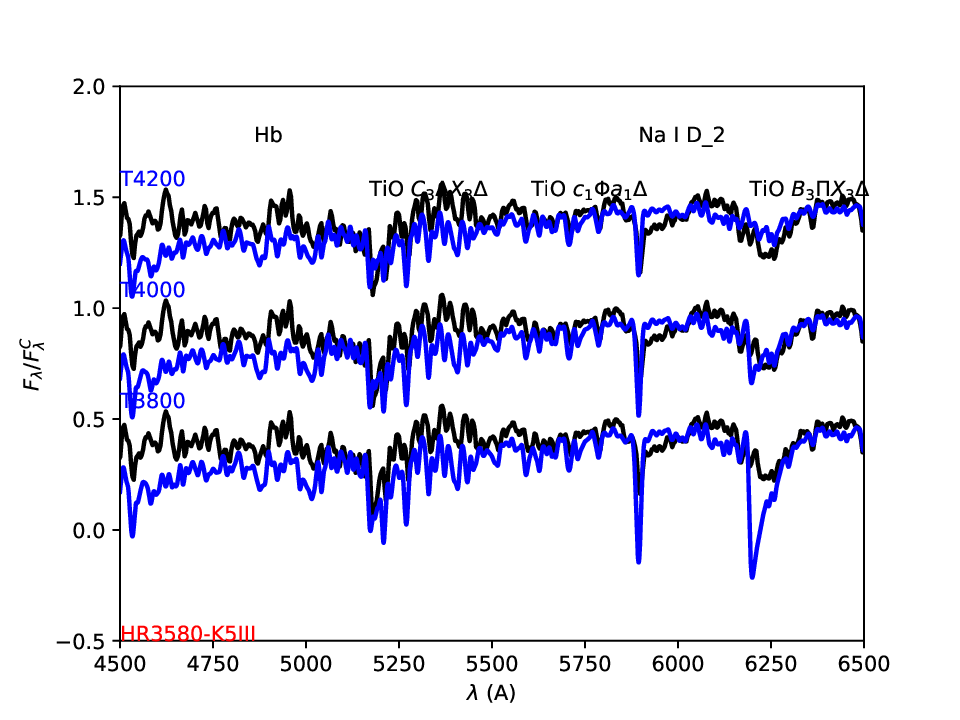}
\caption{HR3580 (K5 III):  Black:  Observed spectrum from our BGO program plotted
with three different vertical offsets.  Blue:
Synthetic spectra for models of $\log g = 2.0$, $[{{\rm A}\over {\rm H}}] = 0.0$, 
and $T_{\rm eff}$ values of 3800, 4000, and 4200 K
in order of increasing vertical offset.
\label{HR3580}
}
\end{figure}

\begin{figure}
\includegraphics[width=\columnwidth]{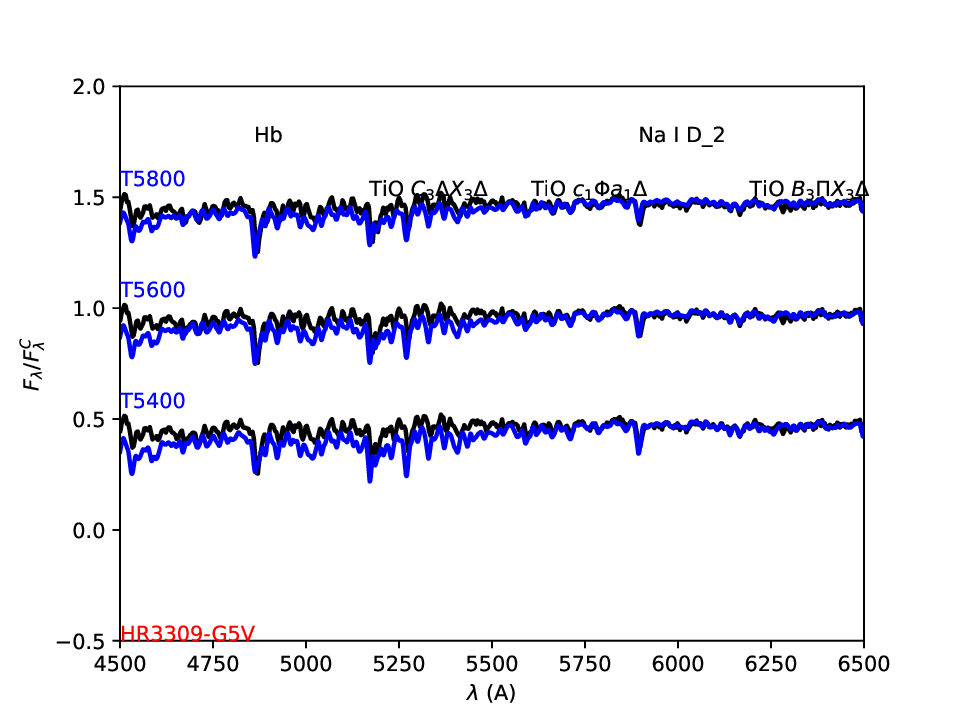}
\caption{Same as Fig. \ref{HR3580} except for star HR3309 (G5 V) and models
of $T_{\rm eff}$ values of 5400, 5600, and 5800 K. 
\label{HR3309}
}
\end{figure}

\begin{figure}
\includegraphics[width=\columnwidth]{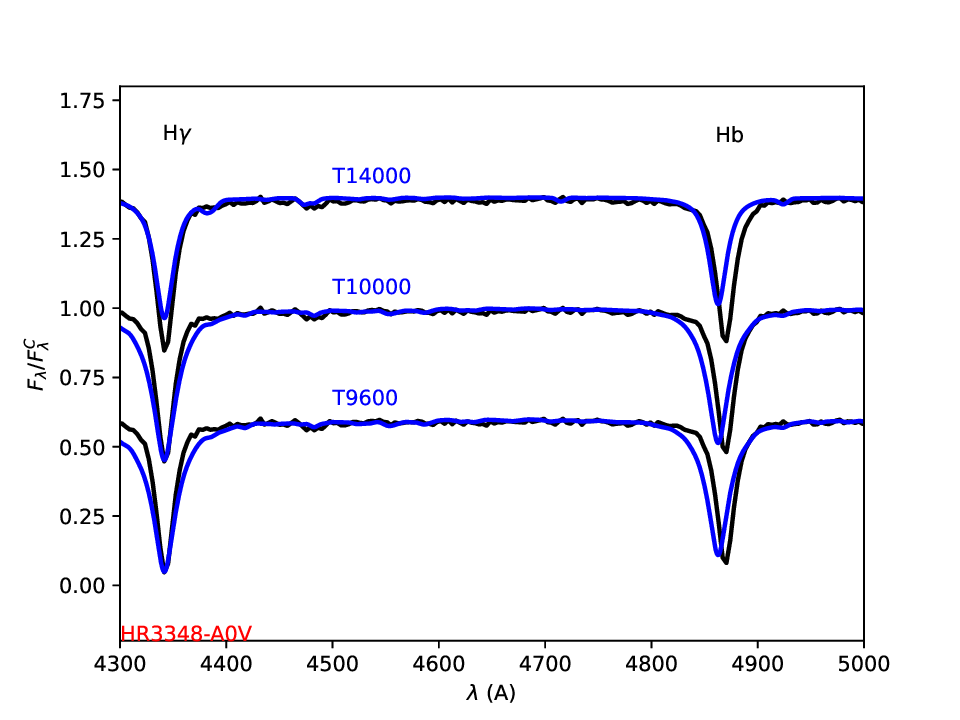}
\caption{Same as Fig. \ref{HR3580} except for star HR3348 (A0 V) and models
of $T_{\rm eff}$ values of 9600, 10000, and 14\, 000 K. 
\label{HR3348}
}
\end{figure}

In a companion paper we describe a grid of stellar atmospheric models computed
with the ChromaStarServer (CSS) code
that spans the 
$T_{\rm eff}$ range 3600 to 22\, 000 K with $\log g = 4.5$ and $[{{\rm A}\over{\rm H}}] = 0.0$,
with $\Delta T_{\rm eff}$ intervals of 200 K for $T_{\rm eff} \le 8000$ K and 400 K for
$T_{\rm eff} > 8000$ K along with models of $\log g = 2.0$ at select $T_{\rm eff}$ values 
less than 5000 K and the corresponding CSS synthetic spectra in the $\lambda$ range
400 to 750 nm.
In Figs. \ref{HR3580}, \ref{HR3309}, and \ref{HR3348} we show the comparison between 
observed spectra from our BGO observing run and synthetic spectra from our model grid 
bracketing the nominal $T_{\rm eff}$ value corresponding to the spectra type listed 
in \citet{hoffleit}.  Nominal $T_{\rm eff}$ values were taken from Appendix G of \cite{appendixG}.
Because we can only achieve an approximate continuum rectification for late type stars 
with high $\Delta\lambda/\Delta x$ broadband data, we do not attempt a quantitative
fit based on minimizing a fitting statistic, but only a perform a visual inspection of the
fit quality.  

\paragraph{}

As discussed in the companion paper, at the low $R$ and high ${{\Delta\lambda}\over{\Delta x}}$ values of these spectra, 
the main features at which
we can assess the fit within the $\sim 4200$ to $\sim 6800$ \AA~ rectification range for GK stars are the
\ion{Na}{1} $D_{\rm 2}$ doublet at $\lambda 5900$ \AA~ and the TiO
C$_{\rm 3}\Delta$-X$_{\rm 3}\Delta$ ($\alpha$ system, $\lambda_{\rm 00}~ 5170$ \AA) and the
B$_{\rm 3}\Pi$-X$_{\rm 3}\Delta$ ($\gamma$' system, $\lambda_{\rm 00}~ 6193$ \AA) bands.
For A and B stars, at our $R$ and ${{\Delta\lambda}\over{\Delta x}}$ values, the main features at which
we can judge the fit within the rectification range are the \ion{H}{1}
$\beta$ and $\gamma$ lines.  This is sufficient to allow students to do projects at the
undergraduate honours level in which they carry out and reduce their own BGO spectroscopy to
coarsely classify stars to within a few spectral subclasses accuracy.  In the process, they
will gain valuable experience with the procedures of observational and computational stellar spectroscopy
within a Python IDE running on commonplace Windows or Linux computer.

\acknowledgements

This work was made possible by the ACENET research computing consortium (ace-net.ca/) and the Digital Research Alliance of Canada (alliancecan.ca). 
This work has made use of the VALD database, operated at Uppsala University, the Institute of Astronomy RAS in Moscow, and the University of Vienna
(vald.astro.uu.se/).  We are also grateful for useful discussion with Brian Skiff of Lowell Observatory.

\end{document}